\documentclass[reprint,showpacs,preprintnumbers,amsmath,amssymb,floatfix]{revtex4-1}
\usepackage{float}
\usepackage{color}
\usepackage{graphicx}
\usepackage{dcolumn}
\usepackage{bm}
\usepackage{xcolor}
\usepackage{comment}
\usepackage[colorlinks,citecolor=blue,linkcolor=red,anchorcolor=blue,filecolor=blue,urlcolor=blue]{hyperref}
\usepackage{color}
\usepackage{csquotes}
\usepackage{comment}
\begin{document}
\title{Role of nuclear deformation and orientation about symmetry axis of target nucleus on heavy-ion fusion dynamics} 

\author{Shilpa Rana$^1$}
\email{srana60\_phd19@thapar.edu}
\author{M. Bhuyan$^{2,3}$}
\email{bunuphy@um.edu.my}
\author{Raj Kumar$^1$}
\email{rajkumar@thapar.edu}
\author{B. V. Carlson$^4$}
\email{brettvc@gmail.com}
\affiliation{$^1$School of Physics and Materials Science, Thapar Institute of Engineering and Technology, Patiala, Punjab 147004, India}
\affiliation{$^2$Institute of Physics, Sachivalaya Marg, Bhubaneswar 751005, Odisha, India} 
\affiliation{$^3$Center for Theoretical and Computational Physics, Department of Physics, Faculty of Science, Universiti Malaya, Kuala Lumpur 50603, Malaysia}
\affiliation{$^4$Institute Tecnologico de Aeroneutica, Sao Jose dos Campos, Sao Paulo, 1222900, Brazil}
\bigskip 
\begin{abstract}
Nuclear shape and orientation degrees of freedom are incorporated into the calculation of the double-folding nuclear potential within the relativistic mean-field (RMF) formalism. The quadrupole deformations ($\beta_2$), nuclear densities and the effective nucleon-nucleon (NN) interaction potential are obtained using the RMF approach for the Hybrid, NL3$^*$ and NL3 parameterizations. The calculated quadrupole deformations are included in the target densities through the nuclear radius. The deformation and orientation-dependent microscopic nuclear potentials are further employed to obtain fusion barrier characteristics and cross-sections for 12 even-even heavy-ion reactions with doubly magic spherical $^{16}$O and $^{48}$Ca as projectiles along with deformed targets from different mass regions. The results obtained for the relativistic R3Y NN potential are compared with those of the Reid version of the non-relativistic M3Y NN potential as well as with the available experimental data. A decrease in the barrier height and increase in the cross-section is observed upon the inclusion of target quadrupole deformations in the nuclear density distributions at the target orientation angles, $\theta_2\le58^\circ$ for the R3Y NN potential and at $\theta_2\le60^\circ$ for the M3Y NN potential. On comparing the $\theta_2$-integrated cross-section calculated using M3Y and R3Y NN potentials with spherical and deformed densities, one observes that the deformed densities and the relativistic R3Y NN potential obtained for the Hybrid parameter set provide better agreement with the available experimental data for all the considered reactions. Moreover, the modifications in the characteristics of the fusion barrier and hence in the cross-section with the inclusion of nuclear shape degrees of freedom and orientations are found to become more prominent in reactions forming heavier compound nuclei. This implies that the inclusion of nuclear deformations and orientation in the calculation of the microscopic nuclear potential within the RMF formalism is crucial to provide a reliable description of the sub-barrier nuclear fusion dynamics, especially in the heavy and superheavy mass regions.
\end{abstract}
\maketitle
\section{INTRODUCTION}
The investigation of heavy-ion fusion reactions in the low-energy regime plays a vital role in exploring various phenomena, ranging from energy generation in stars to the extension of the nuclear chart through the synthesis of new elements. Consequently, a large number of studies have focused on understanding the mechanism of nuclear fusion reactions, both in experimental and theoretical contexts \cite{canto20,mont17,toub17,das98,raj20,jiang21,stef10,back14,wakhle18,bala95,hagino03}. Theoretical investigations of fusion reactions involving heavy ions usually begin with the interpretation of the total potential between the two interacting nuclei \cite{mont17,toub17,canto20,das98,raj20,back14}. The long-range repulsive part of this total potential includes the well-defined Coulomb potential, which depends upon the electric charge, and the centrifugal potential, which depends upon the angular momentum of the fusing nuclei. The short-range nuclear potential contributes an important attractive part to the total interaction potential \cite{raj11,deep13,rajk11,deep14,back14,bloc77,bloc81,cheng19,bass73,bass74,bass77,vaut72,raj20,chamon02,chamon21}. In the literature, numerous macroscopic, semi-classical, and microscopic approaches have been constructed to evaluate this nuclear potential \cite{raj11,deep13,rajk11,deep14,back14,bloc77,bloc81,cheng19,bass73,bass74,bass77,vaut72,raj20,chamon02,chamon21}.

A well-founded technique to obtain the nuclear potential is the double-folding approach \cite{satc79}, in which the nucleus-nucleus potential is determined by integrating the overlap of nuclear densities with an effective nucleon-nucleon (NN) interaction. The Paris \cite{anan83} and Reid \cite{bert77} versions of the Michigan 3 Yukawa (M3Y) effective NN potential have been frequently utilized to calculate this nucleus-nucleus optical potential. Recently, the double-folding approach was applied using nuclear densities and the microscopic R3Y NN interaction developed within the relativistic mean-field (RMF) approach. The nuclear potential thus obtained was implemented successfully in studies of nuclear fusion and cluster radioactivity \cite{chamon21,sing12,sahu14,lahi16,bhuy18,bhuy20,rana21,rana21a,kumar22,bhuy22,rana22,rana22a}. In all these studies, for the sake of simplicity, the nuclear potential was calculated assuming the spherical symmetry of the interacting nuclei. However, nuclear fusion is a complicated phenomenon which may be influenced by numerous factors such as the nuclear shape degrees of freedom and related orientations, shell effects, pairing energy, the mass and isospin asymmetry of the entrance channel \cite{mont17,toub17,canto20,das98,raj20,back14,pengo83,gautam17,tora17,tian09,sime08,stok78,gupta06,gupta04,nishi01,sarg11}. Therefore, including these characteristics in the evaluation of the nuclear interaction potential is essential for a better understanding of the dynamics of complex fusion.

Taking into account these various factors, the influence of nuclear deformations on nuclear reactions and decay dynamics has been studied extensively \cite{mont17,toub17,canto20,das98,raj20,back14}. This is because the shape of nuclei tends to deviate from a spherical one due to the long-range correlations, as one moves away from the magic shell nuclei. The most prominent deformations observed in atomic nuclei are the quadrupole ones, in which the nuclei still maintain reflection and axial symmetries \cite{raman01,prit12,moll16,gorg16,nach04,ibbo98,bes61,tow73,krish68,will66}. Consequently, many theoretical as well as experimental studies have been done to probe the impact of the quadrupole deformations of interacting nuclei on the reaction and decay mechanisms. Furthermore, with the inclusion of degrees of freedom of the nuclear shape in the description of the interaction potential, the relative orientations of the colliding nuclei also affect the characteristics of the nuclear interaction potential \cite{mont17,toub17,canto20,das98,raj20,back14}. In the present study, our aim is to move forward from our previous studies by including the effect of quadrupole deformation and the orientation of the symmetry axis of the target nucleus in microscopic calculations of the nuclear potential within the RMF formalism.

The self-consistent RMF formalism has been used to study infinite and finite nuclear matter characteristics, such as nuclear deformation parameters, charge radii, and binding energies \cite{vret05,meng16,ring96,lala09,lala97,rein86,sing12,sahu14,lahi16,meng06,dutra14,afan05,piek09}. Here, we calculate the quadrupole deformation ($\beta_2$) for each target nucleus using the axially deformed RMF theory with the well-known NL3 \cite{lala97} set and its revised versions, the NL3$^*$ \cite{lala09} and Hybrid \cite{piek09} parameterizations. The influence of quadrupole deformations and related orientations is included in the target densities through the nuclear radius. The nuclear potential is obtained by folding the overlap of the deformed densities with the relativistic R3Y effective nucleon-nucleon (NN) potential and results are also compared with the non-relativistic M3Y NN potential. The validity of this nuclear potential obtained using the RMF formalism along with the inclusion of target deformations and orientations is then assessed by probing the nuclear fusion dynamics of twelve reactions, namely, $^{16}$O+$^{148,150}$Nd, $^{16}$O+$^{154}$Sm, $^{16}$O+$^{176}$Yb, $^{16}$O+$^{176,180}$Hf, $^{16}$O+$^{182,186}$W, $^{48}$Ca+$^{32}$S, $^{48}$Ca+$^{154}$Sm, $^{48}$Ca+$^{168}$Er and $^{48}$Ca+$^{238}$U. These heavy-ion reactions have doubly magic spherical $^{16}$O and $^{48}$Ca projectiles that fuse with deformed target nuclei from different mass regions of the nuclear chart. Fusion and/or capture cross-sections are determined from the $\ell-$summed Wong model \cite{wong73,kuma09} and the theoretical cross-sections are compared with the available experimental data \cite{broda75,leigh95,rajb16,leigh88,mont13,knya07,saga03,nishi12}.

The paper is structured as follows. Sec. \ref{theory} is devoted to a detailed description of the theoretical formalism adopted to calculate the cross-section by incorporating the target deformations and orientations. The results obtained are elaborated in Sec. \ref{rslts}, while Sec. \ref{smry} summarizes and concludes the findings of the present analysis. 

\section{THEORETICAL FORMALISM}
\label{theory}
The total interaction potential between two colliding heavy-ions forms an important part of the foundation for understanding the complex reaction dynamics. As this interaction potential depends upon the structural properties of the colliding nuclei, here we aim to include the impact of nuclear shape degrees of freedom and orientation in the description of the nucleus-nucleus interaction potential. The total interaction potential formed between a deformed target having quadrupole deformation $\beta_2$, and a spherical projectile, can be written as,
\begin{eqnarray}
V_T(R,\beta_2,\theta_2)&=&V_C(R,\beta_2,\theta_2)+V_n(R,\beta_2,\theta_2)+\frac{\hbar^2\ell(\ell+1)}{2\mu R^2}. \nonumber \\
\label{vt}
\end{eqnarray}
Here, $V_C(R,\beta_2,\theta_2)$ and $V_n(R,\beta_2,\theta_2)$ are the deformed, orientation-dependent Coulomb and nuclear potentials, respectively. The inter-nuclear separation distance is denoted by $R$, $\mu$ represents the reduced mass, and $\theta_2$ is the angle between $\vec{R}$ and the axis of symmetry of the deformed target nucleus.  The nuclear potential between a spherical projectile and deformed target nucleus is obtained here using the double folding approach \cite{satc79}, i.e.,
\begin{eqnarray}
V_{n}(\vec{R},\beta_2,\theta_2) &=& \int\rho_{p}(\vec{r}_p)\rho_{t}(\vec{r}_t(\beta_2,\theta_2))\nonumber \\
&& V_{eff}
\left( |\vec{r}_p-\vec{r}_t +\vec{R}| {\equiv}r \right) d^{3}r_pd^{3}r_t, 
\label{vnn}
\end{eqnarray}
where  $\rho_p(\vec{r}_p)$ and $\rho_t (\vec{r}_t (\beta_2,\theta_2))$ denote the total densities (sum of neutron and proton densities) for the spherical projectile and axially deformed target nucleus, respectively. The term $V_{eff}$ in Eq. (\ref{vnn}) is the nucleon-nucleon interaction potential. 

The densities in Eq. \ref{vnn} are calculated self-consistently using the relativistic mean-field (RMF) approach. The RMF formalism has been used frequently to explore the characteristics of finite nuclei lying close and far from the line of $\beta-$stability \cite{vret05,meng16,ring96,lala09,lala97,rein86,sing12,sahu14,lahi16,meng06,dutra14,afan05}. In the RMF approach, the inter-nucleon interaction is mediated via the exchange of photons and mesons. A phenomenological Lagrangian density, which defines the nuclear-meson many-body interactions, can be written as \cite{vret05,meng16,ring96,lala09,lala97,rein86,sing12,sahu14,lahi16,meng06,dutra14,afan05}, 
 \begin{eqnarray}
{\cal L}&=&\overline{\psi}\{i\gamma^{\mu}\partial_{\mu}-M\}\psi +{\frac12}\partial^{\mu}\sigma
\partial_{\mu}\sigma \nonumber \\
&& -{\frac12}m_{\sigma}^{2}\sigma^{2}-{\frac13}g_{2}\sigma^{3} -{\frac14}g_{3}\sigma^{4}
-g_{\sigma}\overline{\psi}\psi\sigma \nonumber \\
&& -{\frac14}\Omega^{\mu\nu}\Omega_{\mu\nu}+{\frac12}m_{\omega}^{2}\omega^{\mu}\omega_{\mu}
-g_{w}\overline\psi\gamma^{\mu}\psi\omega_{\mu} \nonumber \\
&&-{\frac14}\vec{B}^{\mu\nu}.\vec{B}_{\mu\nu}+\frac{1}{2}m_{\rho}^2
\vec{\rho}^{\mu}.\vec{\rho}_{\mu} -g_{\rho}\overline{\psi}\gamma^{\mu}
\vec{\tau}\psi\cdot\vec{\rho}^{\mu}\nonumber \\
&&-{\frac14}F^{\mu\nu}F_{\mu\nu}-e\overline{\psi} \gamma^{\mu}
\frac{\left(1-\tau_{3}\right)}{2}\psi A_{\mu}.
\label{lag}
\end{eqnarray}
The symbol $\psi$ in this equation represents a Dirac nucleon of mass $M$. The terms $m_\sigma$, $m_\omega$, $m_\rho$ and $g_\sigma$, $g_\omega$, $g_\rho$ denote the masses and nucleon-meson coupling constants for the $\sigma$, $\omega$, $\rho$ mesons, respectively. The terms $g_2$ and $g_3$ denote the non-linear self-interactions of the isoscalar scalar $\sigma$-mesons. The meson masses, nucleon-meson and meson self-coupling constants are the principal parameters of the RMF formalism and are often adjusted to match the experimental data for the ground state observables of a selected number of closed-shell nuclei. A number of the RMF parameter sets are available in the literature \cite{dutra14}. Here, the widely adopted NL3 \cite{lala97} parameter set is employed to calculate the quadrupole deformation, nuclear densities and R3Y effective NN potential. Calculations are also performed with the NL3$^*$ parameter set \cite{lala09}, which is a revised version of the set NL3 that produces a value of the incompressibility of nuclear matter (K = 258.28 MeV) within its current acceptable range \cite{garg18,grams22}. Moreover, we have also used the Hybrid parameter set in the present study, which produces a comparatively soft equation of state with K=230.01 MeV \cite{piek09}. The symbols $\tau$ and $\tau_3$ in Eq. (\ref{lag}) denote the nucleon isospin and its third component, respectively. 

The quantities,  $\Omega^{\mu\nu}$,  $\vec B^{\mu\nu}$ and $F^{\mu\nu}$, represent the field tensors for $\omega$, $\rho$-mesons and photons, respectively and are given as,
\begin{eqnarray}
\Omega_{\mu\nu} = \partial_{\mu} \omega_{\nu} - \partial_{\nu} \omega_{\mu}\\
\vec{B}_{\mu \nu}=\partial_{\mu} \vec{\rho}_{\nu} -\partial_{\nu} \vec{\rho}_{\mu}
\end{eqnarray}
and
\begin{eqnarray}
F_{\mu\nu} =\partial_{\mu} A_{\nu}-\partial_{\nu} A_{\mu}.
\end{eqnarray}
Here, $A_\mu$ represents the electromagnetic field. The equations of motions for mesons as well as Dirac nucleons are derived by solving the Euler-Lagrange equations in the mean-field approximation and are written as,  
\begin{eqnarray}
&& \Bigl(-i\alpha.\bigtriangledown+\beta(M+g_{\sigma}\sigma)+g_{\omega}\omega+g_{\rho}{\tau}_3{\rho}_3 \Bigr){\psi} = {\epsilon}{\psi}, \nonumber \\
&& \left( -\bigtriangledown^{2}+m_{\sigma}^{2}\right) \sigma(r)=-g_{\sigma}{\rho}_s(r)-g_2\sigma^2 (r) - g_3 \sigma^3 (r),\nonumber  \\ 
&& \left( -\bigtriangledown^{2}+m_{\omega}^{2}\right) \omega(r)=g_{\omega}{\rho}(r),\nonumber   \\  
&& \left( -\bigtriangledown^{2}+m_{\rho}^{2}\right) \rho(r)=g_{\rho}{\rho}_3(r).
\label{field}
\end{eqnarray} 
The isoscalar scalar $\sigma-$ provides a short-range attractive NN interaction, and the isoscalar vector $\omega-$meson provides a short-range repulsive NN interaction. The isovector vector $\rho-$meson contributes a short-range attractive interaction between protons and neutrons and a short-range repulsive potential between nucleons of identical isospin. This indicates that an effective NN interaction can be obtained by solving the RMF equations for mesons in the single meson exchange limit. We call the effective relativistic NN interaction the R3Y potential \cite{sing12,sahu14,lahi16,bhuy18}. It is given as a function of the inter-nucleon separation (r) as,   
 \begin{eqnarray}
V_{eff}^{R3Y}(r)=\frac{g_{\omega}^{2}}{4{\pi}}\frac{e^{-m_{\omega}r}}{r}
+\frac{g_{\rho}^{2}}{4{\pi}}\frac{e^{-m_{\rho}r}}{r}
-\frac{g_{\sigma}^{2}}{4{\pi}}\frac{e^{-m_{\sigma}r}}{r} \nonumber \\
+\frac{g_{2}^{2}}{4{\pi}} r e^{-2m_{\sigma}r}
+\frac{g_{3}^{2}}{4{\pi}}\frac{e^{-3m_{\sigma}r}}{r}
+J_{00}(E)\delta(r). 
\label{r3y}
\end{eqnarray}
Here, the last term represents a pseudopotential which accounts for the one pion exchange potential (OPEP). The quadrupole deformation ($\beta_2$) for the considered target nuclei is calculated using the RMF formalism on an axially deformed basis. The spherically symmetric radial densities for projectile and target nuclei are also calculated using the RMF Lagrangian defined in Eq. (\ref{lag}). The pairing correlations are taken into account within the well-known Bardeen–Cooper–Schrieffer (BCS) approach \cite{zeng83,moli97,lala99,lala99a,doba84,madl88}.

The nuclear shape degrees of freedom for the target nucleus are included in the description of the nuclear density through the radius vector. The radius of a deformed and oriented target nucleus can be written in terms of a spherical harmonic expansion as \cite{moll16, bohr52,bohr53}, 
\begin{eqnarray}
{r}_t(\beta_2,\theta_2)={r}_{0t}[1+\sqrt{(5/4\pi)}\beta_2 P_2(cos\theta_2)].
\label{drad}
\end{eqnarray}
where, $r_{0t}$ is the corresponding spherical radius.  With the inclusion of the nuclear shape degrees of freedom, the orientation ($\theta_2$) of the symmetry axis of the quadrupole-deformed target relative to the inter-nuclear separation vector ($\vec{R}$) also affects the nuclear interaction potential. This deformation- and orientation-dependent nuclear interaction potential is computed using Eq. (\ref{vnn}), using the R3Y effective NN potential and the nuclear densities and quadrupole deformations obtained within the self-consistent RMF formalism. The results of the relativistic R3Y NN potential are also compared with the Reid version of the non-relativistic M3Y NN potential, which is given in terms of three Yukawa components \cite{satc79,bert77} as,  
\begin{eqnarray}
V_{eff}^{M3Y}(r)= 7999 \frac{e^{-4r}}{4r}-2140\frac{e^{-2.5r}}{2.5r}+J_{00}(E)\delta(r). 
\label{m3y}
\end{eqnarray}
The nuclear potential calculated using the M3Y and R3Y effective NN interactions, along with the Coulomb and centrifugal potentials generates the total interaction potential (see Eq. \ref{vt}). The repulsive Coulomb potential between a spherical projectile and axially deformed target nuclei that have atomic numbers $Z_p$ and $Z_t$, respectively, is given as \cite{wong73},
\begin{eqnarray} 
V_C(R,\beta_2,\theta_2)&=&\frac{Z_p Z_t e^2}{R} \bigg[1+ \frac{r_t^2 \beta_2 P_2(cos\theta_2)}{R^2}\nonumber \\ && \bigg(\sqrt{\frac{9}{20\pi}} +\frac{3\beta_2 P_2(cos\theta_2)}{7\pi}\bigg)\bigg]. 
\label{vc}
\end{eqnarray}
To obtain the Coulomb potential, $r_{0t}$ is taken from the expression $r_{0t}=1.28 A_t^{1/3}-0.76 + 0.8 A_t^{-1/3}$ \cite{bohr52,bohr53}, where, $A_t$ is the atomic mass number of the deformed target nucleus. Here, we have not taken into account additional correction terms in the Coulomb potential, since their contribution is usually very small for the heavy-ion fusion at near-barrier energies \cite{taki00}. Also, the angle between the radius vector and the symmetry axis of the target nucleus is fixed at a given $\theta_2$ such that the separation distance between the surfaces of the interacting nuclei is minimum \cite{zagr04,gupta04,isma10,chop22,isma15}. Moreover, the Coulomb reorientation of deformed nuclei is not taken into account in the present study as its extent is observed to be small for reactions involving deformed targets with heavier mass \cite{sime07,desai11,godre10} and the considered reactions also involve heavier deformed targets except for $^{32}$S which has comparatively smaller $\beta_2$(Hybrid)$=0.147$. The sum of the attractive and repulsive potentials leads to the formation of the fusion barrier. The properties \textit{i.e.} height, position, and curvature of this barrier determine the fusion probability. After determining the total potential (Eq. \ref{vt}) at each target orientation angle ($\theta_2$), the barrier height ($V_B^\ell$) and the barrier position ($R_B^\ell$) are obtained using
\begin{eqnarray}
 \frac {dV_{T}^{\ell}}{dR} \bigg |_{R=R_{B}^{\ell}}=0.
 \label{vb1}
\end{eqnarray}
 and
 \begin{eqnarray}
 \frac{d^2V_{T}^{\ell}}{dR^2}\bigg |_{R=R_{B}^{\ell}}\le 0.
 \label{vb2}
 \end{eqnarray}
Moreover, the barrier curvature ($\hbar\omega_\ell$) is determined as, 
\begin{eqnarray}
\hbar \omega_{\ell}=\hbar [|d^2V_{T}^{\ell}(R)/dR^2|_{R=R_{B}^{\ell}}/\mu]^{\frac{1}{2}}.
\label{vb3}
\end{eqnarray}

The barrier properties are further used to calculate the barrier penetration probability in the parabolic barrier approximation. The Hill-Wheeler transmission coefficient \cite{hill53} through an inverted harmonic oscillator-shaped barrier is widely adopted to obtain the heavy-ion fusion probability \cite{wong73,kuma09,bhuy18,bhuy22} and is given as,  
 \begin{eqnarray}
P_\ell(E_{c.m},\theta_2)=\Bigg[1+exp\bigg(\frac{2 \pi (V_{B}^{\ell}(\theta_2)-E_{c.m.})}{\hbar \omega_{\ell}(\theta_2)}\bigg)\Bigg]^{-1}. 
\end{eqnarray}
Here, $E_{c.m.}$ represents the centre of mass energy of the target-projectile system. Finally, the fusion and/or capture cross-section is calculated at each target orientation angle ($\theta_2$) using the modified version of the simple Wong formula \cite{wong73,kuma09}. This extended version of the Wong formula is known as the $\ell-$summed Wong model and incorporates term-by-term alterations of the interaction potential due to its dependence on the angular momentum \cite{kuma09,bhuy18,bhuy22}. Within the $\ell-$summed Wong model, the cross-section is written in terms of the $\ell-$partial waves as, 
\begin{eqnarray}
\sigma(E_{c.m.},\theta_2)=\frac{\pi}{k^{2}} \sum_{\ell=0}^{\ell_{max}}(2\ell+1)P_\ell(E_{c.m},\theta_2).
\label{crs}
\end{eqnarray}
Here, $k=\sqrt{\frac{2 \mu E_{c.m.}}{\hbar^{2}}}$ and the $\ell_{max}$ values are determined using the sharp cut-off \cite{beck81} model in the energy region above the barrier. In the sub-barrier region, an energy-dependent extrapolation is adopted to obtain the $\ell_{max}$ values. The $\ell-$summed cross-section is obtained using Eq. (\ref{crs}) at each target orientation angle ($\theta_2$ varying from $0^{\circ}$ to $90^{\circ}$ with at step size of $1^{\circ}$) and is further integrated over the target orientation angle ($\theta_2$) to obtain the integrated fusion cross-section, i.e., 
\begin{eqnarray}
\sigma_{int}(E_{c.m.})=\int_{0}^{\pi/2} \sigma(E_{c.m.},\theta_2) sin\theta_2 d\theta_2.
\label{icrs}
\end{eqnarray}
This equation is used to obtain a cross-section for reactions involving a deformed target nucleus fusing with a spherical projectile.

\section{Results and Discussion}
\label{rslts}
The structural properties of two heavy ions play a pivotal role in the understanding of the nuclear fusion mechanism. In this direction, the appropriate nuclear shape degree of freedom and orientations are incorporated in the cross sections through the total interaction potential from the self-consistent relativistic mean-field approach. The RMF formalism has a wide range of applications, from exploring the properties of infinite nuclear matter to understanding the various structural characteristics of finite nuclei \cite{vret05,meng16,ring96,lala09,lala97,rein86,sing12,sahu14,lahi16,meng06,dutra14,afan05}. In our recent studies \cite{bhuy18,bhuy20,rana21,rana21a,kumar22,bhuy22,rana22,rana22a}, spherical nuclear densities and R3Y effective NN interaction potential evaluated using the RMF formalism has also been adopted to explore the nuclear fusion dynamics of heavy-ion reactions. In the present study, we have incorporated the influence of the target quadrupole deformations and orientations in the calculation of the nuclear interaction potential. The deformation and orientation dependent nuclear potential is further employed to obtain the fusion barrier characteristics and cross-section for several heavy-ion reactions using doubly magic spherical and/or nearly spherical $^{16}$ O and $^{48}$ Ca projectiles fusion with deformed rotational target nuclei from different mass regions, namely, $^{32}$S, $^{148,150}$Nd, $^{154}$Sm, $^{168}$Er, $^{176}$Yb, $^{176,180}$Hf,  $^{182,186}$W and $^{238}$U. The calculations are performed using the NL3 and NL3$^*$ parameter sets, which have been found to be suitable for studying nuclear fusion \cite{rana22,bhuy22}. The Hybrid parameter set \cite{piek09} with a relatively soft EoS is also used in the present study.

\begin{figure*}
\centering
\includegraphics[scale=0.35]{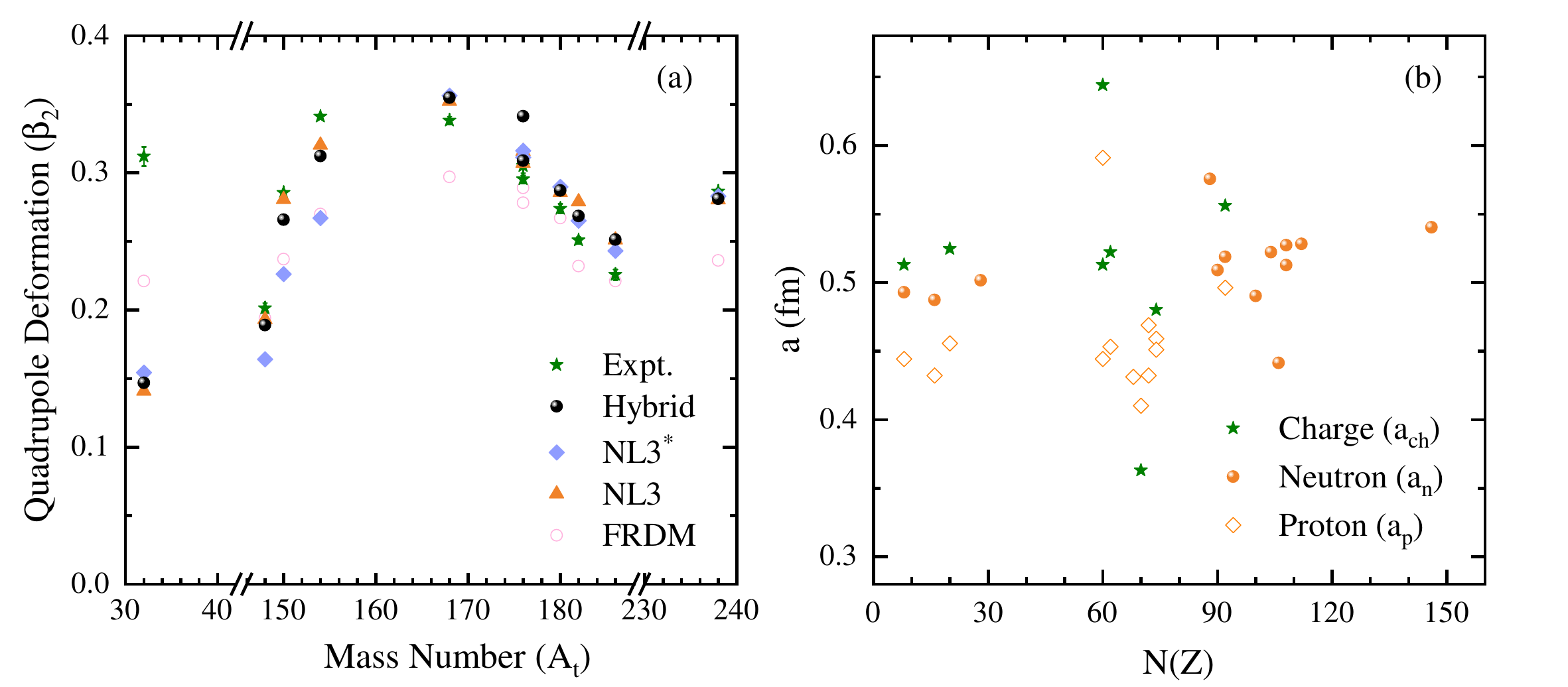}
\caption{ (a) The quadrupole deformation parameter ($\beta_2$) for the target nuclei obtained from axially deformed RMF calculations using the Hybrid (black spheres), NL3$^*$ (blue squares) and  NL3 (orange triangles) parameter sets. The finite range droplet model (FRDM) \cite{moll16} (magenta circles) and the experimental values \cite{raman01,prit12} (green stars) are plotted for comparison. (b) Surface diffusion parameter for all targets and projectiles considered. The experimental data is taken form \cite{vrie87}.}
\label{fig1}
\end{figure*}

To explore the impact of nuclear structure properties on the reaction dynamics, first, the ground state quadrupole deformations ($\beta_2$) are obtained from the RMF formalism in the axially deformed harmonic oscillator basis for all the considered target nuclei. Figure \ref{fig1}(a) depicts the $\beta_2$ values obtained from the Hybrid (black spheres), NL3$^*$ (blue squares) and NL3 (orange triangles) sets,  as a function of target mass number ($A_t$). The predicted $\beta_2$ values from the finite-range droplet model (FRDM) \cite{moll16} (magenta circle) and the experimental data \cite{raman01,prit12} (green stars) are shown in Fig. \ref{fig1}(a) for comparison. A reasonable match can be observed between the $\beta_2$ values obtained from the RMF formalism and the experimental data, as well as with the FRDM predictions, especially in the heavy mass region. The $\beta_2$ values obtained for different RMF sets show a slight deviation in the lighter mass region, whereas, the values match in the heavier mass region. Furthermore, all target nuclei studied have positive $\beta_2$ values, therefore exhibiting a rugby ball-like prolate shape in their ground state. Unlike spherically symmetric nuclei, the radius and density of an axially deformed nucleus are orientation dependent, leading to the dependence of the interaction potential on the orientation angle ($\theta_2$) between the spherical projectile and the deformed target. Thus, it is crucial to take into account the effect of nuclear deformation and the resulting orientation dependence for a better understanding of the nuclear reaction dynamics.

The surface diffuseness is an important parameter of the nuclear density distributions and is often used to study the surface properties of atomic nucleus. The nuclear surface diffuseness parameter is also calculated for the nuclei considered to correlate the surface properties with the nuclear density distributions.  The equivalent nuclear surface diffuseness parameter for the density distributions can be obtained using the relation $a_i \approx -\rho_i/ \frac{d\rho_i}{dr}$, where $i$ stands for the proton- ($a_p$), neutron- ($a_n$) and charge ($a_{ch}$) of the nucleus. Figure \ref{fig1}(b) displays the calculated surface diffuseness parameters for neutron (orange spheres) and proton (orange squares) density distributions from the RMF approach. Here, we have only plotted the surface diffuseness parameter for the NL3 parameter set, since all the considered parameter sets give similar results for the densities and radii of the nuclei considered. Hence, the results for the surface diffuseness are also expected to overlap. The surface diffuseness for the charge density (green stars) from \cite{vrie87} is shown in Fig. \ref{fig1}(b). The magnitude of the surface diffuseness parameter is higher for neutron densities than the proton densities for the considered nuclei. This observation also persists for nuclei with N=Z. Further, the values of the calculated surface diffuseness parameter for proton density are smaller than those for the experimental charge density. This difference of $\approx 0.1$ $fm$ arises due to the finite size effect of the proton density, which is not taken into account in the RMF calculations.  

\begin{figure*}
\centering
\includegraphics[scale=0.27]{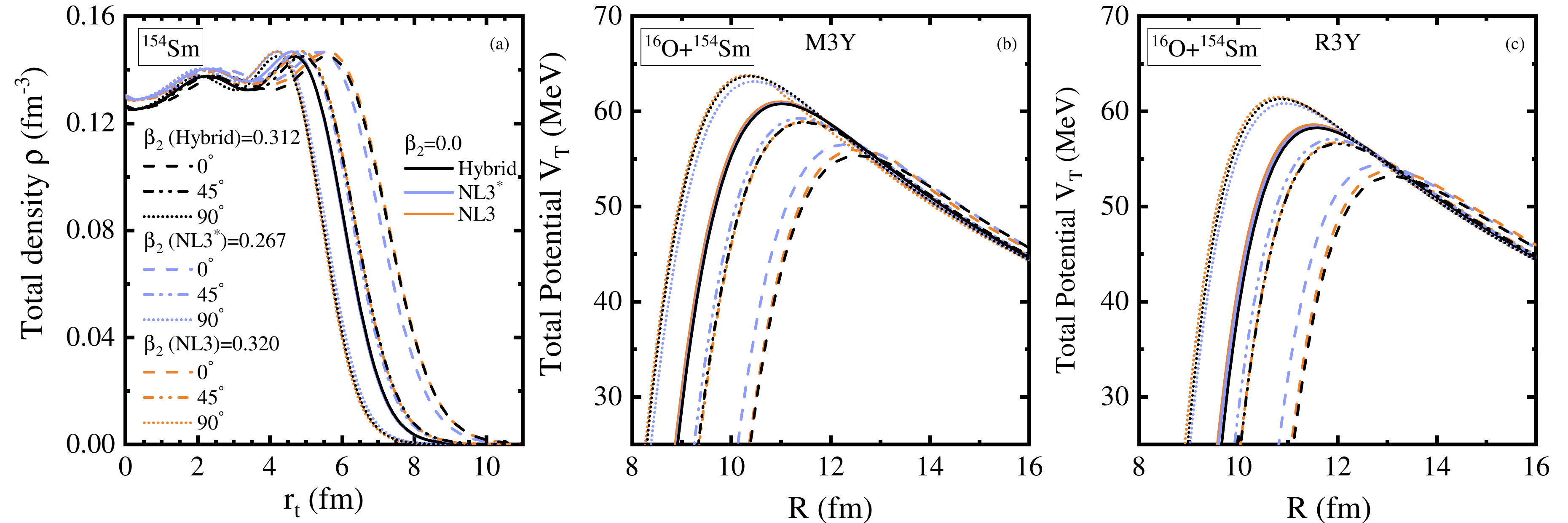}
\caption{(a) Total RMF densities as a function of target radius $r_t$ (fm) for the $^{154}$Sm nucleus for spherical (solid lines) and deformed cases at different orientation angles ($\theta_2$). (b) The total  $s-$wave ($\ell=0$) potential $V_T$ (MeV) as a function of the inter-nucleus separation $R$ (fm) obtained by folding the M3Y NN potential with deformed RMF-NL3 densities plotted for the illustrative case of the $^{16}$O+$^{154}$Sm reaction at different target orientation angles ($\theta_2$). (c) Same as (b) but for the R3Y NN potential.}
\label{fig2}
\end{figure*}

The $\beta_2$ values obtained from the RMF formalism are used to include the nuclear shape degrees of freedom and orientations in the nuclear densities through the nuclear radius [see Eq. (\ref{drad})]. Figure \ref{fig2}(a) shows the deformed RMF total density (sum of neutron and proton densities) obtained for Hybrid (black lines),  NL3$^*$ (blue lines) and NL3 (orange lines) parameter sets at three orientation angles $\theta_2=0^{\circ}$ (dashed lines), $45^{\circ}$ (dash double dotted lines) and $90^{\circ}$ (dotted lines) in the illustrative case of the $^{154}$Sm nucleus. A significant change in the density distribution in the surface region with respect to the spherically symmetric RMF density (solid lines) can be attributed to the inclusion of quadrupole deformations ($\beta_2$). For a given parameter set, the highest surface density is observed at $\theta_2=0^{\circ}$. Further, a decrease in surface density is observed with the increase in the $\theta_2$ and a minimum density is observed at $\theta_2=90^{\circ}$. On comparing the nuclear densities obtained for different parameter sets, it is noted that the spherical density in the surface region increases with the decrease in the nuclear incompressibility (K) of the parameter set, with the Hybrid and NL3 parameter sets giving the highest and lowest densities, respectively. For the case of deformed density, the NL3$^*$ parameter set gives the lowest peripheral density at $\theta_2=0^{\circ}$. This is because the NL3$^*$ parameter set gives the lowest value of $\beta_2$ for the $^{154}$Sm nucleus. However, at $\theta_2=90^{\circ}$, the NL3 parameter set gives the lowest surface density and NL3$^*$ gives the highest density. As nuclear fusion is a surface phenomenon, a small shift in the peripheral density leads to modification of the fusion barrier characteristic \cite{gupta07,rana22}. To explore the effect on nuclear fusion of this change in the target densities in the surface region due to the inclusion of nuclear shape degrees of freedom and orientations, the nuclear interaction potential is obtained by integrating the deformed densities with the R3Y NN potential obtained from the RMF formalism. The results of the microscopic R3Y NN potential are also compared with the widely adopted non-relativistic M3Y NN potential. The repulsive Coulomb potential obtained using Eq. \ref{vc} and the centrifugal potential are added to the nuclear potential to obtain the total interaction potential for the 12 reactions under study.

Figures \ref{fig2}(b) and \ref{fig2}(c) show the values of the total  $s-$wave ($\ell=0$) potential $V_T(R)$ (MeV) versus the inter-nuclear separation $R$ (fm) at orientation angles $\theta_2=0^{\circ}$ (dashed lines), $45^{\circ}$ (dash double dotted lines) and $90^{\circ}$ (dotted lines) calculated using the M3Y and R3Y NN potentials, respectively, for the illustrative case of the $^{16}$O+$^{154}$Sm reaction. The total interaction potentials obtained for the spherically symmetric target density, i.e., $\beta_2=0$ (solid lines) are also shown for comparison. The NL3 and Hybrid sets are observed to give the highest and lowest barrier heights for the case of a spherical target. Noticeable modifications in the height and positions of the fusion barrier are observed with the inclusion of the quadrupole deformation of the target nucleus depending upon the orientation angle ($\theta_2$).  For a given parameter set, the highest fusion barrier and shortest interaction radius (Hot, compact configuration) are observed at $\theta_2=90^{\circ}$, whereas the lowest barrier height and longest interaction radius (Cold, elongated configuration) are obtained at $\theta_2=0^{\circ}$ for both the M3Y and R3Y NN potentials. These results are in line with the optimum orientations \cite{manh05,gupta05} observed for the case of colliding spherical and prolate nuclei using the phenomenological proximity potential. Moreover, at $\theta_2=0^{\circ}$ and $\theta_2=45^{\circ}$, the highest barrier is obtained for NL3$^*$ parameter set, which gives the lowest value of $\beta_2$ for $^{154}$Sm. The Hybrid parameter set gives the lowest fusion barrier. However,  at $\theta_2=90^{\circ}$ the height of the barrier increases with the increase in $\beta_2$ value of the target nucleus. All these observations show that the characteristics of the fusion barrier depend significantly upon the nuclear deformations and orientations as well as on the choice of the EoS. Furthermore, a lower barrier height for the R3Y NN potential when compared to the M3Y NN potential is observed at all orientation angles. This is because the relativistic R3Y NN potential, developed in terms of nucleon-meson couplings (see Eq. \ref{r3y}), furnishes a much more attractive nuclear potential than the non-relativistic Reid M3Y NN potential, written in terms of Yukawa terms that are fit to reproduce the G-matrix elements in an oscillator basis \cite{satc79,bert77}.

\begin{figure*}
\centering
\includegraphics[scale=.1]{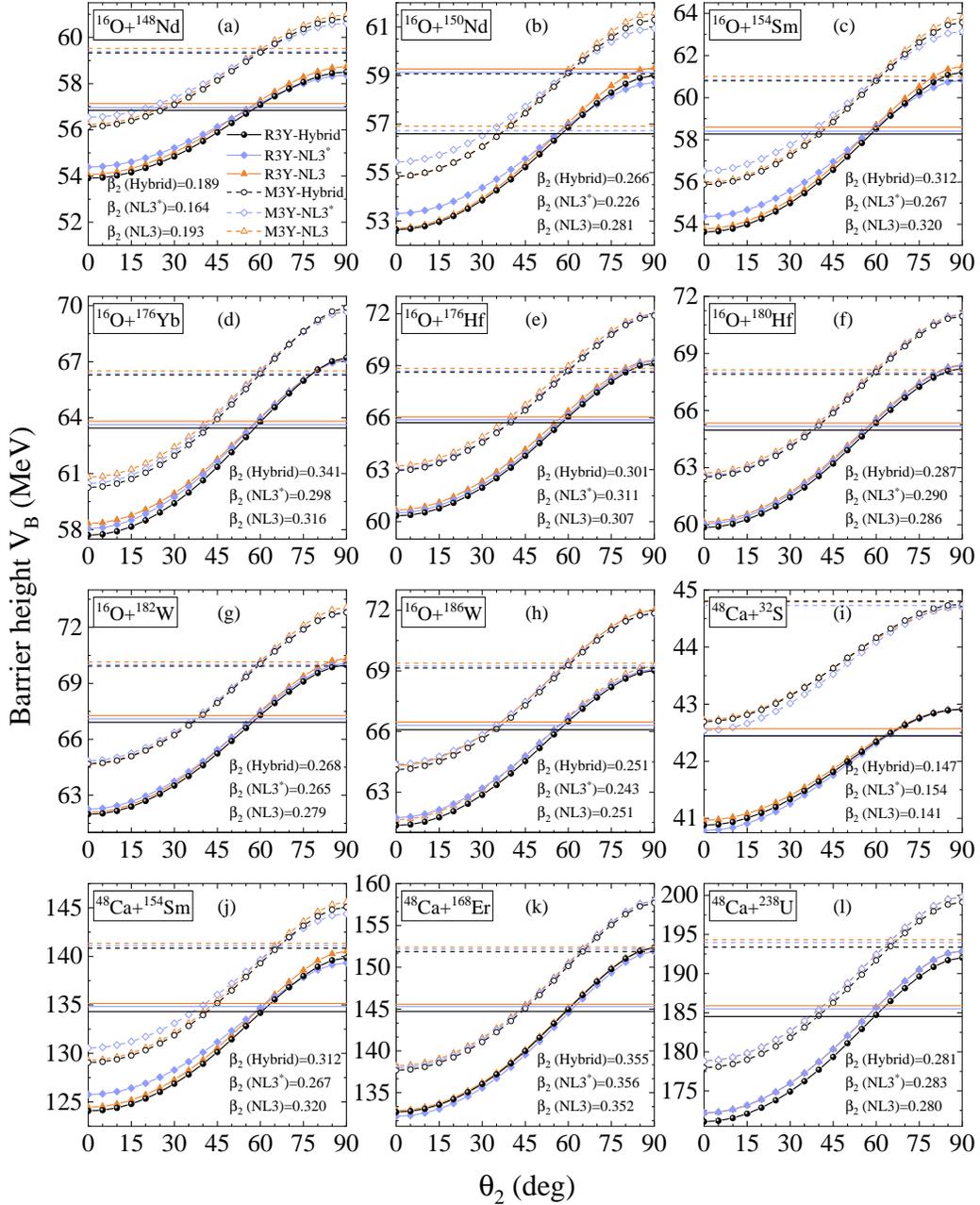}
\caption{ Variation in the height of the $s-$wave ($\ell=0$) barrier $V_B$ (MeV) with target orientation angle ($\theta_2$) for the nuclear potential obtained by folding the M3Y  and R3Y interactions with deformed RMF densities obtained for Hybrid (black), NL3$^*$ (blue) and NL3 (orange) sets.}
\label{fig3}   
\end{figure*}

For a more comprehensive and quantitative investigation of the variation of the fusion barrier height with the target quadrupole deformation and orientations, the $s-$wave ($\ell=0$) barrier heights $V_B$ (MeV) are plotted as a function of target orientation angle ($\theta_2$) in Fig. \ref{fig3} for the 12 reactions under study i.e., (a) $^{16}$O+$^{148}$Nd, (b) $^{16}$O+$^{150}$Nd, (c) $^{16}$O+$^{154}$Sm, (d) $^{16}$O+$^{176}$Yb, (e) $^{16}$O+$^{176}$Hf, (f) $^{16}$O+$^{180}$Hf, (g) $^{16}$O+$^{182}$W (h) $^{16}$O+$^{186}$W, (i) $^{48}$Ca+$^{32}$S, (j) $^{48}$Ca+$^{154}$Sm, (k) $^{48}$Ca+$^{168}$Er and (l) $^{48}$Ca+$^{238}$U. Here, the filled symbols represent the barrier heights obtained by using the R3Y NN potential and deformed RMF densities, whereas, the hollow symbols represent those obtained using the M3Y NN potential integrated with deformed RMF densities. The solid and dashed lines without symbols in Fig. \ref{fig3} denote the barrier height obtained using spherically symmetric RMF densities along with the R3Y and M3Y NN potentials, respectively. The quadrupole deformation parameters ($\beta_2$) calculated from the axially deformed RMF  approach using the different parametrizations for the target nuclei are also denoted in the respective panels of Fig. \ref{fig3}. It can be clearly noted from the panels of Fig. \ref{fig3} that the R3Y NN potential gives a lower barrier than the M3Y NN potential in the case of a spherical density, for all the reactions under study. This observation also persists at a given $\theta_2$, when the effect of target quadrupole deformation is included. Also, the lowest and highest barrier height are observed at $\theta_2=0^{\circ}$ and $\theta_2=90^{\circ}$, respectively, for both the M3Y and R3Y NN potentials for all the reactions under study. This is because the target nuclei are all prolate ($\beta_2>0$) in their ground state and when a prolate target collides with a spherical projectile, the interaction radius will be maximum (Cold, elongated configuration) at $\theta =0^\circ$ and minimum (Hot, compact configuration) at $\theta =90^\circ$. The increase in the barrier height on changing $\theta_2$ from $0^\circ$ to $90^\circ$ is smallest (1.94 MeV for R3Y and 1.98 MeV for M3Y) for the $^{48}$Ca+$^{32}$S reactions, in which the lightest compound nucleus (CN) is formed and is the largest (20.72 MeV for R3Y and 20.97 MeV for M3Y) for $^{48}$Ca+$^{238}$U, leading to the formation of a CN in the superheavy mass region. For the $^{48}$Ca+$^{168}$Er reaction, involving the target nucleus with the highest $\beta_2=0.352$, the change in barrier height on changing $\theta_2$ from $0^\circ$ to $90^\circ$ is 19.51 (19.76) MeV for the R3Y (M3Y) NN potential. Moreover, this shift in the barrier height is slightly smaller for the R3Y NN potential in comparison to the M3Y NN potential for the reactions considered. On comparing the barrier heights obtained including target quadrupole deformation with those obtained using spherical RMF densities, we observe that the former are smaller than the latter at $\theta_2 \le 60^\circ$  for the M3Y NN potential and at $\theta_2 \le 58^\circ$ for the R3Y NN potential for all the considered reactions. The $\theta_2$ value at which the barrier height obtained for the deformed case surpasses that obtained for the spherical case is always smaller for the R3Y NN potential than the M3Y NN potential for all the reactions. Moreover, this value is smaller ($\theta_2 \ge 58^\circ$) for $^{16}$O induced reactions than that ($\theta_2 \ge 61^\circ$) for $^{48}$Ca induced reactions.

On comparing the barrier heights obtained for different RMF parameter sets for the case of s spherical target, it can be noted that the Hybrid parameter set gives the lowest barrier, whereas the NL3 set gives the highest fusion barrier for all the reactions under study. Thus, it can be inferred that the parameter set with the softest EoS gives the lowest barrier height. This trend also persists at a given $\theta_2$ when the effect of target quadrupole deformation is included, but only for the reactions for which the Hybrid, NL3$^*$ and NL3 parameter sets give similar values of $\beta_2$. Moreover, the barrier height at higher orientation angles is directly proportional to the quadrupole deformation, whereas at lower orientation angles, the trend of the barrier height obtained for different parameterizations depends upon both the $\beta_2$ values and choice of EoS. A more careful study of Fig. \ref{fig3} shows that the change in the barrier height upon the inclusion of quadrupole deformation is smaller for the R3Y NN potential than the M3Y interaction at smaller orientation angles. This trend however becomes inverted at $\theta_2\ge59^\circ$ for the $^{16}$O induced reactions and at $\theta_2\ge63^\circ$ for the $^{48}$Ca induced reactions. The barrier heights obtained by folding the M3Y NN potential with quadrupole deformed densities are smaller than those obtained by folding the R3Y NN potential with spherical RMF densities at $\theta_2\le 30^\circ$. The barrier height obtained for the R3Y NN potential folded with deformed RMF densities surpasses the barrier height obtained for the M3Y NN potential folded with spherical RMF densities at larger orientation angles ($\theta_2\ge 87^\circ$) for the $^{16}$O induced reactions involving target nuclei having $\beta_2\ge 0.279$ i.e., $^{16}$O+$^{150}$Nd, $^{16}$O+$^{154}$Sm, $^{16}$O+$^{176}$Yb, $^{16}$O+$^{176,180}$Hf and $^{16}$O+$^{182}$W. All these observations imply that the inclusion of target quadrupole deformations and the related orientation significantly modify the properties, such as the height, position and shape, of the total interaction potential generated between two fusing nuclei.

\begin{figure*}
\centering
\includegraphics[scale=0.25]{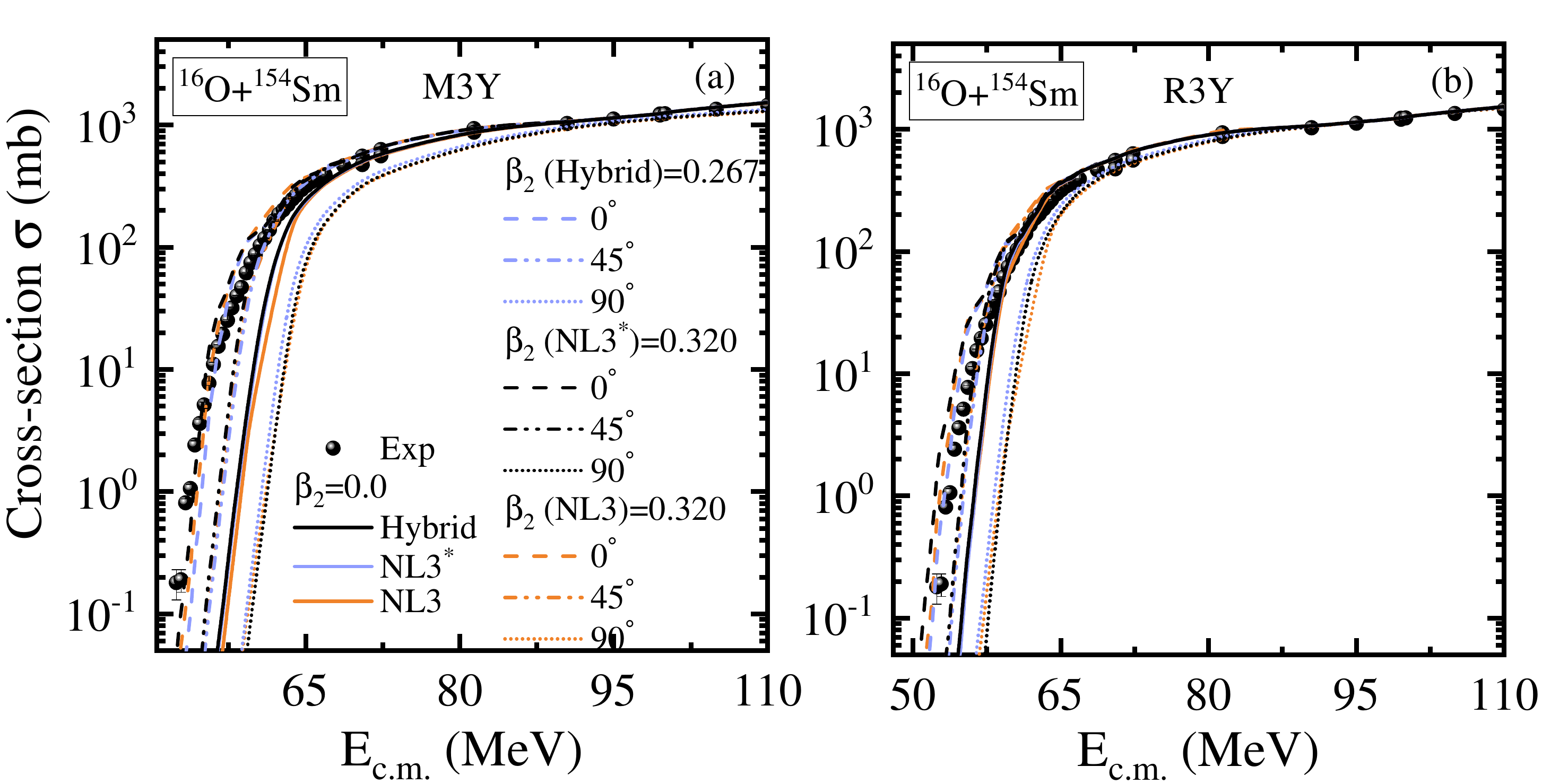}
\caption{The cross-section $\sigma$ (mb) as a function of the center-of-mass energy $E_{c.m.}$ (MeV) obtained using (a) the M3Y and (b) the R3Y NN potentials and deformed densities for the $^{16}$O+$^{154}$Sm reaction at different target orientation angles ($\theta_2$).}
\label{fig4}
\end{figure*}

The properties of the interaction barrier are prerequisites to obtaining the fusion probability for two interacting heavy-ions. Here, we have used the Hill-Wheeler approximation \cite{hill53} to obtain the transmission coefficient, which is further employed to calculate the cross-section from the $\ell-$summed Wong model \cite{wong73,kuma09,bhuy18,bhuy22}. In recent studies, the $\ell-$summed Wong model equipped with nuclear potential calculated using spherical RMF densities and the relativistic R3Y NN interaction has been used frequently to explore fusion dynamics \cite{bhuy18,bhuy20,rana21,rana21a,kumar22,bhuy22,rana22,rana22a}. In the present analysis, we move a step further by including the impact of the nuclear shape degrees of freedom in the calculation of nuclear potentials and cross-sections for the systems $^{16}$O+$^{148,150}$Nd, $^{16}$O+$^{154}$Sm, $^{16}$O+$^{176}$Yb, $^{16}$O+$^{176,180}$Hf, $^{16}$O+$^{182,186}$W, $^{48}$Ca+$^{32}$S, $^{48}$Ca+$^{154}$Sm, $^{48}$Ca+$^{168}$Er and $^{48}$Ca+$^{238}$U reactions. The $\ell_{max}$ values in the energy region above the barrier are obtained from the sharp cut-off model \cite{beck81}, while an energy-dependent extrapolation is used for below-barrier energies. As discussed above, the characteristics of the total interaction potential depend upon the angle between the inter-nuclear separation axis and the axis of symmetry of the quadrupole-deformed target nucleus ($\theta_2$).

To further explore the impact of target orientation on nuclear fusion, the cross-section obtained for Hybrid (black lines), NL3$^*$ (blue lines) and  NL3 (orange lines) sets with the inclusion of target quadrupole deformation is plotted at orientation angles $\theta_2=0^{\circ}$ (dashed lines), $45^{\circ}$ (dash double dotted lines) and $90^{\circ}$ (dotted lines) in Fig. \ref{fig4} for the illustrative case of the $^{16}$O+$^{154}$Sm reaction. Figures \ref{fig4}(a) and \ref{fig4}(b) represent the cross-section obtained using deformed RMF densities as well as the M3Y and R3Y NN potentials, respectively. The cross-section calculated using the spherically symmetric RMF densities (solid lines) and experimental data \cite{leigh95} (black spheres) are also shown here for comparison. It can be clearly seen from Fig. \ref{fig4} that the cross-section changes significantly at near and sub-barrier energies ($E_{c.m.}$) with the inclusion of nuclear shape degrees of freedom i.e. target quadrupole deformation. Further, the cross-section is observed to decrease as the target orientation angle ($\theta_2$) increases, with a minimum cross-section obtained at $\theta_2=90^{\circ}$. This is because the surface density of $^{154}$Sm having a prolate shape ($\beta_2=0.320$) has a minimum radius at $\theta_2=90^{\circ}$ (see Fig. \ref{fig2}(a)), which leads to a higher fusion barrier and a lower cross-section at this orientation. On comparing the cross-section obtained with spherical RMF densities to that obtained using deformed RMF densities, an increase in the cross-section is noted on the inclusion of target quadrupole deformation at $\theta_2\le 45^{\circ}$. Moreover, the Hybrid and NL3 parameter sets are observed to give the highest and lowest cross-section for the case of spherical targets. On the other hand the NL3$^*$ parameter set is observed to give the lowest cross-section at $\theta_2 =0^{\circ}$ and $\theta_2 =45^{\circ}$, whereas the Hybrid parameter set yields the lowest cross-section at $\theta_2 =90^{\circ}$. Further, a higher cross-section is observed using the R3Y NN potential in comparison to the M3Y NN one at each orientation angle. The reason is that the R3Y NN interaction generates a more attractive nuclear potential which leads to a lower fusion barrier and consequently a higher cross-section in comparison to the non-relativistic M3Y NN potential. On comparing the theoretical results with the experimental data, one finds that both the M3Y and R3Y NN potentials underestimate the cross-section when folded with the spherical RMF densities. However, a reasonable correspondence is observed with the inclusion of target quadrupole deformation at $\theta_2=45^{\circ}$ for the R3Y and at $\theta_2= 0^{\circ}$ for the M3Y NN potential. However, during the experimental measurement of a cross-section, the target nuclei are not aligned at a particular angle. So, for a more comprehensive analysis, integrated cross-sections (see Eq. \ref{icrs}) are calculated for the considered heavy-ion reactions.

\begin{figure*}
\centering
\includegraphics[scale=.2]{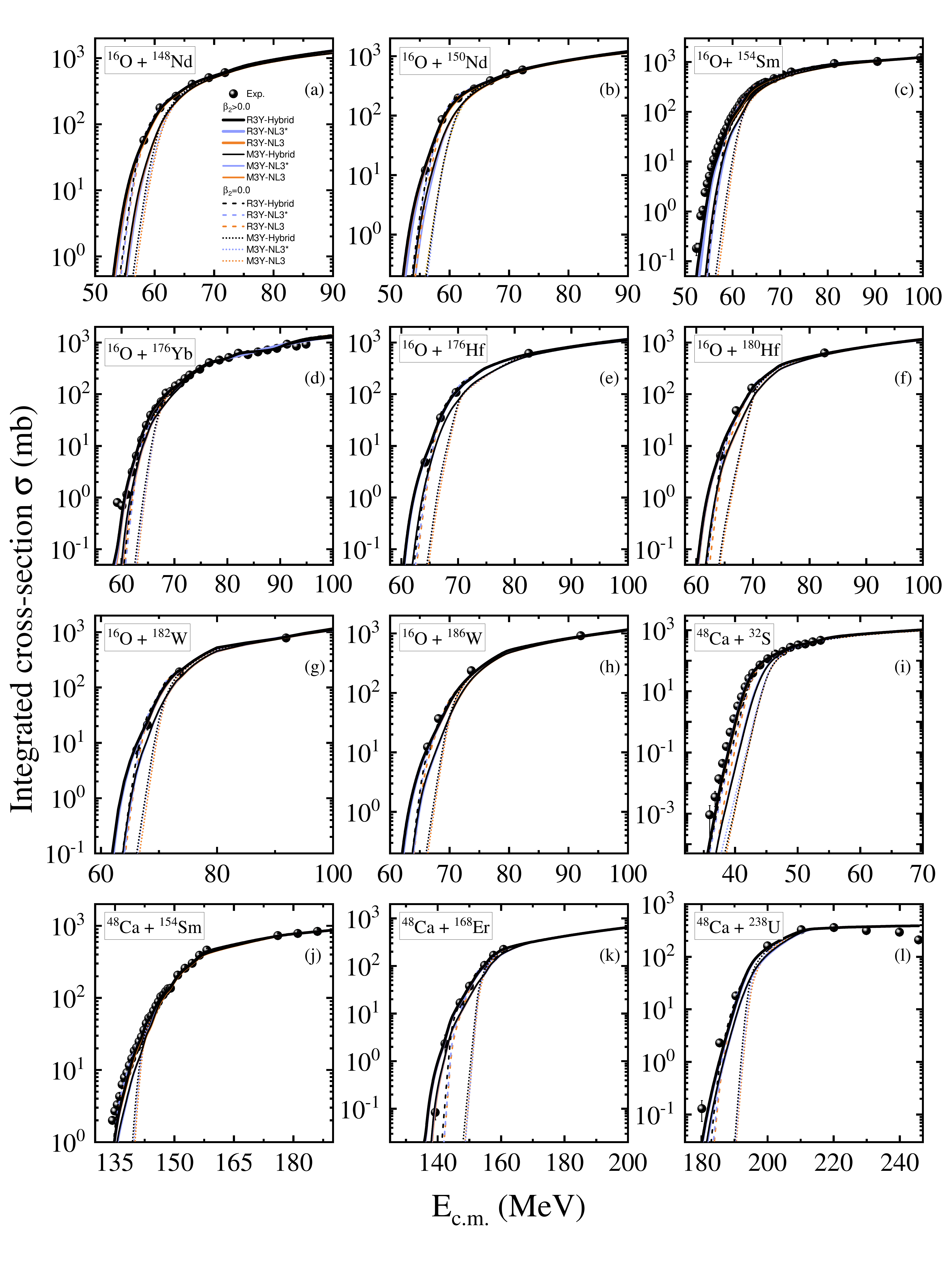}
\caption{The total integrated cross-section $\sigma_{int}$ (mb) obtained using the R3Y and M3Y  NN potentials for spherical and prolate target nuclei. The experimental data \cite{broda75,leigh95,rajb16,leigh88,mont13,knya07,saga03,nishi12} is also given for comparison. See text for details.}
\label{fig5}   
\end{figure*}

Figure \ref{fig5} displays the total $\theta_2-$integrated cross-section $\sigma_{int}$ (mb) calculated using the deformed densities along with the relativistic R3Y (thick solid lines) and M3Y (thin solid lines) effective NN potentials versus the center-of-mass energy $E_{c.m.}$ (MeV) for (a) $^{16}$O+$^{148}$Nd, (b) $^{16}$O+$^{150}$Nd, (c) $^{16}$O+$^{154}$Sm, (d) $^{16}$O+$^{176}$Yb, (e) $^{16}$O+$^{176}$Hf, (f) $^{16}$O+$^{180}$Hf, (g) $^{16}$O+$^{182}$W (h) $^{16}$O+$^{186}$W, (i) $^{48}$Ca+$^{32}$S, (j) $^{48}$Ca+$^{154}$Sm, (k) $^{48}$Ca+$^{168}$Er and (l) $^{48}$Ca+$^{238}$U reactions. The cross-sections calculated using the spherical RMF densities along with the R3Y (dashed lines) and the M3Y (dotted lines) as well as the experimental data \cite{broda75,leigh95,rajb16,leigh88,mont13,knya07,saga03,nishi12} (black spheres) are also plotted in Fig. \ref{fig5} for comparison. We note that similar $\ell_{max}$-values are found for a given center of mass energy for the 12 different nuclear potentials obtained by folding the M3Y and R3Y NN potentials with spherical and deformed RMF densities obtained for the Hybrid (black lines), NL3$^*$ (blue lines) and  NL3 (orange lines) sets. The Hybrid parameter set, with comparatively soft EoS, is noted to give the highest cross-section, which also shows a better match with the experimental data. A significant enhancement in the cross-section, which becomes more prominent in the sub-barrier energy region can be clearly noticed for all the reactions upon the inclusion of target quadrupole deformation in the description of the nuclear interaction potential within the RMF formalism.  The $\theta_2-$integrated cross-section obtained for spherically symmetric RMF density folded with both M3Y and R3Y interactions is found to underestimate the experimental data for all the considered reactions. This underestimation of the cross-section is smaller for the relativistic R3Y interaction as compared to its non-relativistic M3Y counterpart. The overlap between the experimental data and cross-section calculated from the Reid version of the M3Y NN potential however becomes better on folding with the deformed RMF densities, but still underestimates the experimental data at sub-barrier energies. On the other hand, a reasonable overlap with the experimental data is noted for the cross-section calculated using the microscopic R3Y NN potential and the deformed RMF densities obtained using the considered three non-linear parameterizations for the 12 heavy-ion fusion reactions under study. Moreover, the $\theta_2-$integrated cross-sections calculated by employing different nuclear potentials overlap at above barrier energies because the impact of the deformed nuclear structure is suppressed in the above barrier region and the centrifugal potential plays a major role. All these observations lead to the conclusion that the nuclear potential evaluated from the relativistic R3Y NN potential and RMF densities along with the inclusion of nuclear shape degrees of freedom and orientation is necessary to study nuclear fusion. However, a slight underestimation of the experimental cross-section can be noticed for $^{16}$O+$^{154}$Sm, $^{48}$Ca+$^{32}$S, $^{48}$Ca+$^{154}$Sm and $^{48}$Ca+$^{238}$U reactions at deep sub-barrier energies. This discrepancy might be due to higher order deformations such as the hexadecapole ($\beta_4$) deformation of the target nuclei which has not been incorporated in this analysis. The investigation of the impact of these higher-order deformations and other nuclear structure effects on the fusion and decay dynamics will be pursued in future studies.

\section{Summary and Conclusions}
\label{smry}
The impact of nuclear shape degrees of freedom and orientation on the fusion mechanism is explored using the well-established relativistic mean-field (RMF) approach. First, the quadrupole deformation parameter ($\beta_2$) for the target nuclei considered is calculated within the RMF formalism in an axially deformed harmonic oscillator basis using the non-linear Hybrid, NL3$^*$ and NL3 parameter sets. These $\beta_2$-values for the prolate target nuclei are used to include the effect of nuclear quadrupole deformations and orientation in the RMF density distributions through the nuclear radius. The equivalent surface diffuseness parameters are also calculated for the proton and neutron densities from the RMF formalism. The deformed densities and relativistic R3Y effective NN potential are employed to evaluate the deformation and orientation-dependent nuclear potential using the double folding model. This microscopic nuclear potential is further used to explore the fusion dynamics of twelve even-even heavy-ion reactions namely, $^{16}$O+$^{148,150}$Nd, $^{16}$O+$^{154}$Sm, $^{16}$O+$^{176}$Yb, $^{16}$O+$^{176,180}$Hf, $^{16}$O+$^{182,186}$W, $^{48}$Ca+$^{32}$S, $^{48}$Ca+$^{154}$Sm, $^{48}$Ca+$^{168}$Er and $^{48}$Ca+$^{238}$U reactions with spherical and/or nearly spherical projectile nuclei incident on deformed target nuclei from different mass regions. The results of the relativistic R3Y NN potential are also compared with the Reid version of the well-adopted non-relativistic M3Y NN potential.

The height of the fusion barrier is observed to decrease upon the inclusion of target quadrupole deformations for $\theta_2 \le 58^\circ$  for R3Y NN potential and at $\theta_2 \le 60^\circ$ for R3Y NN potential. The height of the fusion barrier is observed to increase with the increase in $\theta_2$ and the highest barrier and shortest interaction radius are observed at $\theta_2=90^\circ$. The change in the barrier characteristics with respect to the target orientation angle $\theta_2$ becomes more prominent for the formation of compound nuclei in the heavier mass region. The R3Y NN potential is observed to give lower barrier heights at a given $\theta_2$ than the M3Y NN potential for all the reactions under study. The deformation and orientation-dependent fusion barrier properties obtained employing the RMF approach are further used to calculate the cross-section within the $\ell$-summed Wong model. Analogous to nuclear densities in the tail region and the total interaction potential, significant modifications in the cross-section are observed with the inclusion of nuclear shape degrees of degrees of freedom and orientation. The largest and smallest cross-sections are obtained at $\theta_2=0^\circ$ and $\theta_2=90^\circ$, respectively, for both R3Y and M3Y NN potentials folded with quadrupole deformed RMF densities. On comparing the barrier characteristics obtained for different RMF parameter sets, lower barrier height and higher cross-section are noted for the softer EoS. Moreover, the barrier height (cross-section) at higher orientation angles is observed to increase (decrease) with the increase in quadrupole deformation. On the other hand, the barrier characteristics and cross-section at lower orientation angles depend upon both the $\beta_2$ values and choice of EoS. Further, $\theta_2-$integrated cross-sections are obtained for all the considered reactions using the quadrupole deformed RMF densities and the results are compared with those obtained using spherical RMF densities and the available experimental data. An increase in cross-section is observed at around and sub-barrier regions upon the inclusion of target quadrupole deformation effects in the description of the nuclear density distributions, which also leads to better agreement with the experimental data. The nuclear potential obtained using the deformed densities and microscopic R3Y effective NN potential evaluated employing the RMF approach with the non-linear Hybrid parameterization is found to give a better match with the experimental cross-sections when compared to the Reid version of the M3Y NN potential for all the reactions studied here. A more comprehensive analysis of nuclear fusion, with the incorporation of the nuclear shape degrees of freedom of both the reaction partners, i.e. target as well as projectile nuclei, will be carried out in our future studies.    
 
\section*{Acknowledgements}
This work has been supported by the Science and Engineering Research Board (SERB), DST, India File No. CRG/2021/001229 and Ramanujan Fellowship File No. RJF/2022/000140. BVC acknowledges support from grant 2017/05660-0 of the São Paulo Research Foundation (FAPESP), grant 303131/2021-7 of the CNPq and the INCT-FNA project 464898/2014-5.


\end{document}